\documentclass{Interspeech2024}
\usepackage{glossaries,siunitx,pgfplots,multirow,balance,paralist}

\usepackage{cite}
\usepackage{booktabs}
\usepackage{etoolbox,siunitx}
\robustify\bfseries
\usepackage{bbm}

\usepackage{url}
\usepackage{hyperref}
\usepackage[hyphenbreaks]{breakurl}

\interspeechcameraready

\title{Sound Event Bounding Boxes}

\name{Janek}{Ebbers}
\name{François G.}{Germain}
\name{Gordon}{Wichern}
\name{Jonathan}{Le Roux}

\address{Mitsubishi Electric Research Laboratories, Cambridge, MA, USA}
\email{\{ebbers,germain,wichern,leroux\}@merl.com}


\def\thineq{\hspace{-.1em}=\hspace{-.1em}}

\newacronym{MIL}{MIL}{multiple instance learning}
\newacronym{PSDS}{PSDS}{polyphonic sound detection score}
\newacronym{noPSDS}{noPSDS}{non-oracle PSDS}
\newacronym[longplural={sound event bounding boxes}]{SEBB}{SEBB}{sound event bounding box}
\newacronym{SED}{SED}{sound event detection}
\newacronym{TP}{TP}{true positive}
\newacronym{FP}{FP}{false positive}
\newacronym{FN}{FN}{false negative}
\newacronym{DTC}{DTC}{detection tolerance criterion}
\newacronym{GTC}{GTC}{ground-truth intersection criterion}

\begin{document}

\setlength{\abovedisplayskip}{4pt}
\setlength{\belowdisplayskip}{4pt}
\setlength{\abovedisplayshortskip}{2pt}
\setlength{\belowdisplayshortskip}{2pt}
\setlength{\textfloatsep}{5pt}
\maketitle

\begin{abstract}
    Sound event detection is the task of recognizing sounds and determining their extent (onset/offset times) within an audio clip. Existing systems commonly predict sound presence confidence in short time frames. Then, thresholding produces binary frame-level presence decisions, with the extent of individual events determined by merging consecutive positive frames. In this paper, we show that frame-level thresholding degrades the prediction of the event extent by coupling it with the system's sound presence confidence. We propose to decouple the prediction of event extent and confidence by introducing SEBBs, which format each sound event prediction as a tuple of a class type, extent, and overall confidence. %
    We also propose a change-detection-based algorithm to convert legacy frame-level outputs into SEBBs. We find the algorithm significantly improves the performance of DCASE 2023 Challenge systems, boosting the state of the art from $.644$ to $.686$ PSDS1.
\end{abstract}

\vspace{-.1cm}
\section{Introduction}
\vspace{-.1cm}
\glsresetall

Automatically recognizing and processing sounds in diverse environments is a highly desired technology for many applications, such as wildlife monitoring, autonomous driving, and surveillance.
Different sound recognition tasks focus on parsing acoustic scenes at different levels of detail.
In particular, audio tagging ~\cite{virtanen2018computational} and \gls{SED}~\cite{virtanen2018computational,mesaros2021sound} tasks aim at exhaustively inventorying the sounds in a scene.
\gls{SED} differs from audio tagging by requiring identification of the temporal extent of sound events on top of their event class. In mathematical terms, it asks to detect the events $e_j$ ($j\thineq 1,\ldots, J$) expressed as triplets $(c_j,t_{\text{on},j},t_{\text{off},j})$ with $c_j$ the event class label, and $t_{\text{on},j}$ (resp.\ $t_{\text{off},j}$) the event's onset (resp.\ offset) time.

While a few event-level models directly outputting event triplet predictions $\hat{e}_j=(\hat{c}_j,\hat{t}_{\text{on},j},\hat{t}_{\text{off},j})$ have been proposed~\cite{ye2021sound,bhosale2023diffsed}, the large majority of \gls{SED} systems rely on frame-level event presence detection models. As such, thresholding of the presence confidence cannot directly output event predictions, and post-processing is needed to consolidate frame-level event presence predictions into event predictions~\cite{mesaros2021sound}.
In that respect, current state-of-the-art models~\cite{nam2022frequency,Kim2023,xin2023improving,shao2023fine} overwhelmingly compute event predictions as blocks of consecutive frame-level presence predictions (i.e., confidences falling above the aforementioned threshold).
As traditionally understood in detection tasks, the threshold then controls the minimum presence confidence triggering an event detection in a binary fashion. As such, appropriate threshold value(s) can be chosen depending on application requirements, with, for example, some applications requiring high recall and others high precision.
Crucially, the current approach means varying the threshold also affects the event predictions in non-trivial and, we argue, detrimental ways. For example, additional frame-level detections due to a lower threshold can change the detected onset/offset times of a predicted event, or even merge multiple predicted events into a single one. 
This, in turn, substantially diminishes the interpretability of current evaluation procedures.

Here, we show this behavior to be substantially sub-optimal. As a remedy, we propose a new structure for \gls{SED} systems to explicitly decouple the prediction mechanisms for onset/offset times and event presence, by introducing the \glspl{SEBB} output format.
Motivated by bounding box predictions in image object detection~\cite{redmon2016yolo}, the \gls{SEBB} format corresponds to a series of event-level candidates with each a predicted class, onset/offset times and a (scalar) presence confidence.  %
Then, predicted events become a series of \glspl{SEBB} whose presence confidence exceeds a (now event-level) 
confidence threshold. Crucially, this threshold now intuitively controls only whether a \gls{SEBB} is predicted as an event, without affecting its onset/offset times, and eliminates the undesirable behaviors observed with the current approach.

\glspl{SEBB} can be predicted in various ways, including in an end-to-end manner. However, we acknowledge that one reason for the enduring popularity of frame-level models is that \gls{MIL} techniques~\cite{kumar2016audio,shah2018closer,wang2019comparison} allow for training without ground-truth onset and offset times (i.e., \textit{weakly labeled training}), while end-to-end prediction usually requires strongly labeled training data~\cite{ye2021sound}. In that context, we also propose a post-processing algorithm to convert the frame-level presence confidence scores into \glspl{SEBB} for any frame-level system. In it, conversion relies primarily on a change-detection approach. Note that change-/slope-based algorithms can be found in prior post-processing signal chains for \gls{SED} in \cite{cances2019evaluation}. However, these would perform \gls{SED} solely based on change/slope without considering absolute confidence at all and did not ultimately improve performance.
In contrast, we find that deploying our proposed post-processing, on top of unlocking the conceptual benefits of \glspl{SEBB}, substantially improves performance. In particular, our post-processing boosts performance of all 13 considered systems from the recent DCASE 2023 Challenge Task 4a~\cite{dcase2023task4a} and establishes a new state of the art. Source code is publicly available\footnote{\url{https://github.com/merlresearch/sebbs}}.

\vspace{-.1cm}
\section{SED with Sound Event Bounding Boxes}
\subsection{Preliminaries}
\vspace{-.1cm}
As stated earlier, \gls{SED} systems commonly consist of a frame-level multi-label classifier followed by a post-processing to output predicted events. In mathematical terms, the classifier corresponds to the operation ${\mathbf{Y}}={f(\mathbf{X})}$, with $f$ denoting a prediction model, ${\mathbf{X}}={[\mathbf{x}_0, \dots, \mathbf{x}_{N-1}]}$ a sequence of input feature vectors $\mathbf{x}_n$ (e.g., log-mel spectrogram frames), and ${\mathbf{Y}}={[\mathbf{y}_0, \dots, \mathbf{y}_{N-1}]}$ a sequence of frame-level class probability vectors $\mathbf{y}_n$, with $n$ the frame index. $y_{n,c}\in[0,1]$ then represents the predicted confidence of sound class $c$ being present in frame $n$. Note that, in practice, input and output sequence lengths may differ, for example when $f$ is a convolutional neural network with striding and/or pooling. For conciseness, we assume same sequence lengths with no loss of generality. 

$\mathbf{y}_n$ is then fed to post-processing. It may be first (optionally) altered, e.g., by median filtering $y_{n,c}$ in times. Ultimately, event predictions are obtained through a frame-level thresholding operation, turning $y_{n,c}$ into a binary $z_{n,c}={\mathbbm{1}_{[y_{n,c} > \lambda_c]}}$ (with $\lambda_c$ a class-dependent threshold), followed by a merging operation where each block of consecutive $z_{n,c}=1$ into a single detected event $\hat{e}_j$. The detected onset (resp. offset) time then corresponds to the beginning (resp. end) of the first (resp. last) frame of that block.

For applications seeking for meaningful connected event predictions, event-based evaluation is employed, which is recently favored by benchmarks and challenges.
For given sets of predicted and ground truth events, counts of \gls{TP}, \gls{FP}, and \gls{FN} events are obtained, with two main approaches currently in use.
Collar-based evaluation~\cite{mesaros2016metrics} makes determinations based on whether the onset and offset times of a predicted event match the onset and offset times of a ground-truth event of the same class up to a maximal allowed divergence.

Notably, as threshold selection criteria vary widely depending on the target application, the community currently relies heavily on threshold-independent metrics derived from ROC curves to aggregate performance over various thresholds $\lambda_c$ as a single score. For example, the recent DCASE 2023 Task 4 challenge used \gls{PSDS}~\cite{bilen2020framework,ebbers2022threshold} which is computed as the normalized area under the PSD-ROC curve, i.e., the average of class-level ROC curves from intersection-based \gls{TP}/\gls{FP}/\gls{FN} results, plus a penalty on inter-class standard deviation.

Crucially, a known problem resulting from the aforementioned conversion of event predictions centered around frame-level confidence thresholding is that, typically, it leads to \gls{TP}/\gls{FP}/\gls{FN} results for which the ROC curve is no longer monotonic, unlike what is expected in traditional classification. As a workaround monotonicity is restored by only taking the (oracle) best-case operating points into account~\cite{bilen2020framework}, but it results in artificially inflated scores and limits the intuitive interpretation of the metric as a performance score.

\vspace{-.1cm}
\subsection{Effects of Frame-level Thresholding}
\vspace{-.1cm}

\begin{figure}[t]
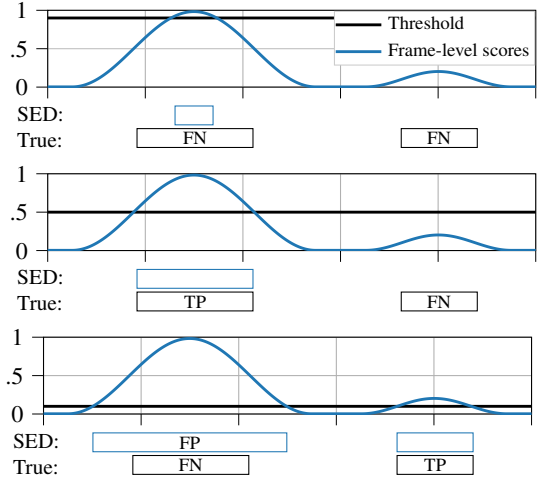

    \centering
    \newlength\figureheight
    \newlength\figurewidth
    \setlength\figureheight{2.6cm}
    \setlength\figurewidth{8cm}
    \input{figures/frame_threshold_09.tex}\\
    \input{figures/frame_threshold_05.tex}\\
    \hspace{0.5mm}\input{figures/frame_threshold_01.tex}
    \vspace{-.25cm}
    \caption{Example of detection with different frame-level thresholds and comparison with ground-truth events.}
    \vspace{-.0cm}
    \label{fig:frame_threshold_example}
\end{figure}

In this section, we show a practical illustration of the impact of frame-level thresholding on event boundary detection, and its detrimental effect on intersection-based evaluation.
For this, we consider the example in Fig.~\ref{fig:frame_threshold_example} as representative of the frame-level presence confidence output for a single sound class found in existing \gls{SED} systems.
We then see how a lower frame-level threshold, while triggering more individual event detections, also has a non-trivial impact on event boundary predictions.

For example, we consider a typical intersection-based evaluation based on the ground-truth events. We use a required intersection rate of $\rho_\text{DTC}=\rho_\text{GTC}=0.7$ for both the \gls{DTC} and \gls{GTC}, i.e., predictions must intersect with a ground-truth event by at least \SI{70}{\%} to not be \gls{FP} and ground-truth events must be covered by detections by at least \SI{70}{\%} to be \gls{TP}~\cite{ronchini2022benchmark}.

Then, we can see that, when gradually lowering the threshold down from $1$, we will first get a prediction corresponding to the first ground-truth event, but with an underestimated extent, leading to \gls{FN}. When lowering the threshold further, that matching prediction remains, but its predicted extent grows longer to the point where it yields \gls{TP}. However, when lowering the threshold even further, the predicted extent will ultimately grow overestimated yielding now both \gls{FN} and \gls{FP}, even as we might get a \gls{TP} in predicting the second ground-truth event.
\glspl{TP} turning back to \glspl{FN} (i.e., having the true positive rate decrease) when the threshold decreases is different from standard binary classification tasks and ultimately makes ROC curves decrease again after some point, breaking their monotonic properties.
As we can see, this is ultimately because the threshold that detects the correct extent depends on the geometry of the frame-level scores (e.g., the overall peak heights in the case of Fig.~\ref{fig:frame_threshold_example}). Crucially, we see that no threshold could get both ground-truth events right at the same time in our example.

This demonstrates how frame-level thresholding is sub-optimal for event detection due to the event-level entanglement of both boundary and confidence information in the frame-level scores. Therefore, we propose to decouple extent and confidence prediction as presented in the next section. 

\begin{figure}[t]
    \centering
    \setlength\figureheight{3cm}
    \setlength\figurewidth{8cm}
    \input{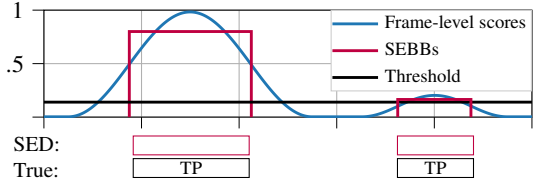}
    \vspace{-2mm}
    \caption{Examples of \glspl{SEBB} with event-level decision threshold and comparison with ground-truth events.}
    \label{fig:sebbs_example}
\end{figure}

\vspace{-.1cm}
\subsection{Sound Event Bounding Boxes}
\vspace{-.1cm}
To solve above issues, we propose the concept of \glspl{SEBB} as new \gls{SED} system output format. In mathematical terms, we define \glspl{SEBB} as quadruples ${\hat{b}_j=(\hat{c}_j,\hat{t}_{\text{on},j},\hat{t}_{\text{off},j},\overline{y}_j)}$ which intuitively represents a sound event candidate defined by sound class $\hat{c}_j$, a fix extent given by onset time $\hat{t}_{\text{on},j}$ and offset time $\hat{t}_{\text{off},j}$, plus an overall presence confidence score $\overline{y}_j$. Fig.~\ref{fig:sebbs_example} shows a graphical representation of \glspl{SEBB} for our earlier example in Fig.~\ref{fig:frame_threshold_example}.
The key idea is that the temporal extent of sound event candidates should be determined independently from the event candidate confidence score.
An event-level thresholding can then be employed to control a system's sensitivity without affecting the temporal extents of event predictions.
In particular, even if the decision threshold is lowered far below a \gls{SEBB}'s confidence score, the temporal extent will not change.
This ensures not to disturb high-confidence event detections when using low decision thresholds, such as in applications aiming for a high recall.
With \glspl{SEBB}, monotonically-increasing ROC curves are thus guaranteed again, and sound event candidates of high and low confidence, as in the above example, may be jointly detected correctly.

\vspace{-.1cm}
\section{SEBBs from Frame-level Outputs}
\vspace{-.1cm}

\label{ssec:SEBBalgos}
Now, as already stated, the vast majority of existing systems outputs frame-level multi-label presence confidence scores. As such, we now present a few post-processing approaches to enable conversion (and thus, evaluation) of their output as inferred \glspl{SEBB}.

\noindent {\bf Threshold-based \texorpdfstring{\glspl{SEBB}}{SEBBs}:}
A simple approach to generate \gls{SEBB} predictions is akin to the threshold-based process illustrated in Fig.~\ref{fig:frame_threshold_example}. Here too, we use median filtering plus frame-level thresholding followed by merging. But, instead of interpreting the results as event predictions, we use the resulting set of $\hat{c}_j,\hat{t}_{\text{on},j},\hat{t}_{\text{off},j}$ together with $\overline{y}_j$, computed as the average over frame-level presence confidence $y_{n,c_j}$ between $\hat{t}_{\text{on},j}$ and $\hat{t}_{\text{off},j}$, as predicted \textit{tSEBB}. As only SEBB selection is to be performed at inference, the thresholds $\lambda_{c,\text{ext}}$ are set jointly with median filter lengths through tuning on a validation set as is commonly done for any frame-level post-processing of $y_{n,c}$~\cite{ebbers2021self,Barahona2023}. A second threshold would be used at inference to turn tSEBBs into predicted events. However, Fig.~\ref{fig:frame_threshold_example} hints at how tSEBBs could still lead to poor detection performance in typical scenarios.

\begin{figure}[t]
    \centering
    \setlength\figureheight{2.92cm}
    \setlength\figurewidth{8cm}
    \hspace{0.3mm}\begin{tikzpicture}

\definecolor{darkgray176}{RGB}{176,176,176}
\definecolor{darkorange25512714}{RGB}{255,127,14}
\definecolor{green01270}{RGB}{0,127,0}
\definecolor{steelblue31119180}{RGB}{31,119,180}
\definecolor{darkviolet1910191}{RGB}{191,0,191}
\definecolor{lightgray204}{RGB}{204,204,204}

\begin{axis}[
height=\figureheight,
width=\figurewidth,
tick align=outside,
tick pos=left,
legend cell align={left},
legend style={
    fill opacity=0.8,
    draw opacity=1,
    text opacity=1,
    at={(0.99,0.46)},
    anchor=south east,
    draw=lightgray204,
    inner sep=0.2pt
},
x grid style={darkgray176},
xmajorgrids,
xmin=-0.03, xmax=8.0,
xtick style={color=black},
xticklabels=\empty,
ytick={0,.5,1},
yticklabels={0,.5,1},
y grid style={darkgray176},
ymajorgrids,
ymin=0, ymax=1,
ytick style={color=black}
]
\addplot [line width=1.08pt, steelblue31119180]
table {%
0.02 0.0539510957896709
0.06 0.0341095775365829
0.1 0.0400215983390808
0.14 0.0349623002111911
0.18 0.0252424906939268
0.22 0.0262306816875934
0.26 0.0345827527344226
0.3 0.0601793453097343
0.34 0.114147000014782
0.38 0.160392016172409
0.42 0.20301941037178
0.46 0.295923411846161
0.5 0.535634577274323
0.54 0.931290745735168
0.58 0.941058456897736
0.62 0.906619369983673
0.66 0.826890826225281
0.7 0.72448593378067
0.74 0.423548728227615
0.78 0.192039996385574
0.82 0.10168644785881
0.86 0.21385857462883
0.9 0.117065876722336
0.94 0.0582655593752861
0.98 0.054598968476057
1.02 0.103405214846134
1.06 0.33791920542717
1.1 0.742905557155609
1.14 0.564411997795105
1.18 0.603452563285828
1.22 0.72919762134552
1.26 0.752803385257721
1.3 0.653361558914185
1.34 0.703738033771515
1.38 0.800517618656158
1.42 0.763020873069763
1.46 0.646031618118286
1.5 0.639431059360504
1.54 0.637493848800659
1.58 0.726470768451691
1.62 0.85011237859726
1.66 0.859433233737946
1.7 0.789640963077545
1.74 0.696695268154144
1.78 0.433563947677612
1.82 0.251943022012711
1.86 0.147407934069633
1.9 0.0978026911616325
1.94 0.0781386271119117
1.98 0.0726712122559547
2.02 0.0701251924037933
2.06 0.0751753896474838
2.1 0.110414743423462
2.14 0.206957638263702
2.18 0.277357846498489
2.22 0.192914351820946
2.26 0.124372988939285
2.3 0.189479663968086
2.34 0.332251518964767
2.38 0.193928107619286
2.42 0.128230363130569
2.46 0.1426160633564
2.5 0.158832430839538
2.54 0.0710952654480934
2.58 0.0459745004773139
2.62 0.0341689139604568
2.66 0.0240453984588384
2.7 0.0187034029513597
2.74 0.0184565708041191
2.78 0.0176068469882011
2.82 0.0168996639549732
2.86 0.015885766595602
2.9 0.0163371171802282
2.94 0.0178659670054912
2.98 0.0192126967012882
3.02 0.0172429308295249
3.06 0.0173713266849517
3.1 0.0191431380808353
3.14 0.0184092726558446
3.18 0.0197944641113281
3.22 0.0230400059372186
3.26 0.0299648102372884
3.3 0.0305204801261425
3.34 0.0312514975666999
3.38 0.0259323008358478
3.42 0.0245834626257419
3.46 0.027126632630825
3.5 0.0265558641403913
3.54 0.0280343051999807
3.58 0.0371732152998447
3.62 0.0576962530612945
3.66 0.139175310730934
3.7 0.33851757645607
3.74 0.786789953708649
3.78 0.931051969528198
3.82 0.874751031398773
3.86 0.664873063564301
3.9 0.32748618721962
3.94 0.211952462792396
3.98 0.10937076061964
4.02 0.0407512448728084
4.06 0.032477606087923
4.1 0.0264414399862289
4.14 0.0195966809988021
4.18 0.0192744992673397
4.22 0.0182482153177261
4.26 0.018196603283286
4.3 0.0175902675837278
4.34 0.0189205538481473
4.38 0.0148359332233667
4.42 0.0117270136252045
4.46 0.010636712424457
4.5 0.0102802403271198
4.54 0.0100093642249703
4.58 0.0090619064867496
4.62 0.0090410271659493
4.66 0.0097102327272295
4.7 0.00981708150357
4.74 0.0092263100668787
4.78 0.0098159927874803
4.82 0.0106908725574612
4.86 0.0113241327926516
4.9 0.0114371450617909
4.94 0.0152307236567139
4.98 0.0244313236325979
5.02 0.0242595095187425
5.06 0.017839528620243
5.1 0.0162642188370227
5.14 0.0157572496682405
5.18 0.0132199609652161
5.22 0.0114816641435027
5.26 0.0115891816094517
5.3 0.0112249935045838
5.34 0.0099403411149978
5.38 0.0091448789462447
5.42 0.0077407592907547
5.46 0.0077103972434997
5.5 0.0082622654736042
5.54 0.0083708371967077
5.58 0.0090773263946175
5.62 0.016431275755167
5.66 0.0125891743227839
5.7 0.0101573765277862
5.74 0.0101768989115953
5.78 0.0101118711754679
5.82 0.0107111250981688
5.86 0.0099575035274028
5.9 0.0093033742159605
5.94 0.0094676790758967
5.98 0.0101281506940722
6.02 0.0101242121309041
6.06 0.010540279559791
6.1 0.0108352089300751
6.14 0.0105620836839079
6.18 0.0105708576738834
6.22 0.0098321866244077
6.26 0.009869473055005
6.3 0.0109469518065452
6.34 0.0108916470780968
6.38 0.0104508139193058
6.42 0.0107718063518404
6.46 0.0126536907628178
6.5 0.0162171591073274
6.54 0.0232847239822149
6.58 0.0183304324746131
6.62 0.0154187586158514
6.66 0.0128878625109791
6.7 0.0121675906702876
6.74 0.0119145270437002
6.78 0.0117677459493279
6.82 0.011992810294032
6.86 0.0113481134176254
6.9 0.011470628902316
6.94 0.0115110017359256
6.98 0.012906152755022
7.02 0.0123874237760901
7.06 0.0116954781115055
7.1 0.0131049547344446
7.14 0.0144565589725971
7.18 0.0130267962813377
7.22 0.0120946979150176
7.26 0.0114143406972289
7.3 0.0112366853281855
7.34 0.0107397455722093
7.38 0.0104146711528301
7.42 0.0109339235350489
7.46 0.0104184169322252
7.5 0.0100421672686934
7.54 0.0106906713917851
7.58 0.0118077946826815
7.62 0.0106796724721789
7.66 0.0098235365003347
7.7 0.0092059383168816
7.74 0.0101885395124554
7.78 0.0106468545272946
7.82 0.0102240992709994
7.86 0.0097253611311316
7.9 0.0107819736003875
7.94 0.0126010458916425
7.98 0.0145059088245034
8.02 0.0219709649682045
8.06 0.0139800161123275
8.1 0.0156827177852392
8.14 0.0148447044193744
8.18 0.0123863164335489
8.22 0.0120648853480815
8.26 0.0112032834440469
8.3 0.0106276739388704
8.34 0.0106574669480323
8.38 0.0108607914298772
8.42 0.010212143883109
8.46 0.0099701099097728
8.5 0.0096621736884117
8.54 0.0098617458716034
8.58 0.0087475376203656
8.62 0.0087405536323785
8.66 0.0091637624427676
8.7 0.0084923226386308
8.74 0.0078307753428816
8.78 0.0064454763196408
8.82 0.0059449397958815
8.86 0.006127817556262
8.9 0.0061311754398047
8.94 0.0053619863465428
8.98 0.0055907997302711
9.02 0.0060552973300218
9.06 0.0061620878987014
9.1 0.0055183782242238
9.14 0.0057368143461644
9.18 0.006150048226118
9.22 0.0060175177641212
9.26 0.0055752811022102
9.3 0.0059167402796447
9.34 0.0059879226610064
9.38 0.005988567136228
9.42 0.0056608528830111
9.46 0.0054961564019322
9.5 0.0057522282004356
9.54 0.0055477879941463
9.58 0.0053180688992142
9.62 0.005368688609451
9.66 0.0057046087458729
9.7 0.0056880745105445
9.74 0.0049073100090026
9.78 0.00548374094069
9.82 0.0063615599647164
9.86 0.0063706990331411
9.9 0.0070070358924567
9.94 0.0083193099126219
9.98 0.0193326380103826
};
\addlegendentry{{\scriptsize Frame-level scores}}
\addplot [line width=1.08pt, purple]
table {%
0 0
0 0.0440297862907983
0.08 0.0440297862907983
0.08 0
};
\addlegendentry{{\scriptsize cSEBBs}}
\addplot [line width=1.08pt, purple]
table {%
0.48 0
0.48 0.755644249574032
0.76 0.755644249574032
0.76 0
};
\addplot [line width=1.08pt, purple]
table {%
1.08 0
1.08 0.699570266998354
1.8 0.699570266998354
1.8 0
};
\addplot [line width=1.08pt, purple]
table {%
2.12 0
2.12 0.194693610606081
2.52 0.194693610606081
2.52 0
};
\addplot [line width=1.08pt, purple]
table {%
3.24 0
3.24 0.0305786744877562
3.36 0.0305786744877562
3.36 0
};
\addplot [line width=1.08pt, purple]
table {%
3.68 0
3.68 0.59077249647937
3.96 0.59077249647937
3.96 0
};
\addplot [line width=1.08pt, purple]
table {%
4.32 0
4.32 0.0109062141863381
9.52 0.0109062141863381
9.52 0
};
\addplot [line width=1.08pt, purple]
table {%
9.76 0
9.76 0.00881246057374901
10 0.00881246057374901
10 0
};
\end{axis}

\end{tikzpicture}\\
    \vspace{-2mm}\hspace{.05mm}
    \begin{tikzpicture}

\definecolor{darkgray176}{RGB}{176,176,176}
\definecolor{darkorange25512714}{RGB}{255,127,14}
\definecolor{green01270}{RGB}{0,127,0}
\definecolor{steelblue31119180}{RGB}{31,119,180}
\definecolor{lightgray204}{RGB}{204,204,204}

\begin{axis}[
height=\figureheight,
width=\figurewidth,
tick align=outside,
tick pos=left,
legend cell align={left},
legend style={
    fill opacity=0.8,
    draw opacity=1,
    text opacity=1,
    at={(0.99,0.71)},
    anchor=south east,
    draw=lightgray204,
    inner sep=0.2pt
},
x grid style={darkgray176},
xmajorgrids,
xmin=-0.03, xmax=8.0,
xtick style={color=black},
xticklabels=\empty,
yticklabels={,-.5,0,.5,},
y grid style={darkgray176},
ymajorgrids,
ymin=-0.8, ymax=0.8,
ytick style={color=black}
]
\addplot [line width=1.08pt, darkorange25512714]
table {%
0 0.035752957376341
0.04 0.0235330509021878
0.08 0.0221930826082825
0.12 0.0278770498310526
0.16 0.042954952456057
0.2 0.0683773572867115
0.24 0.108954365365207
0.28 0.195691059964399
0.32 0.336531332073112
0.36 0.461995674607654
0.4 0.565461947582662
0.44 0.639811030278603
0.48 0.666289328907927
0.52 0.564099716643492
0.56 0.295706025014321
0.6 0.0179921140273412
0.64 -0.22183924416701
0.68 -0.444121971726418
0.72 -0.626585787783067
0.76 -0.669396439567208
0.8 -0.560960444932183
0.84 -0.381692983830969
0.88 -0.178058354184031
0.92 0.0148034909119209
0.96 0.216704720631242
1 0.390629456068079
1.04 0.513634948059916
1.08 0.526836547379692
1.12 0.43213412972788
1.16 0.396927379692594
1.2 0.33265759733816
1.24 0.20636348798871
1.28 0.0792350719372432
1.32 0.0240167280038198
1.36 0.0343334376811982
1.4 0.00324829419453945
1.44 -0.00727769732475277
1.48 0.0305181940396627
1.52 0.0589576164881388
1.56 0.0276139179865519
1.6 -0.0552628288666407
1.64 -0.180646029611429
1.68 -0.323653180152178
1.72 -0.46617179363966
1.76 -0.579719837754965
1.8 -0.606304646780094
1.84 -0.556677961101135
1.88 -0.445726085454225
1.92 -0.300595170507828
1.96 -0.148808244615793
2 -0.0247637120385964
2.04 0.0448507132629554
2.08 0.0933626977105936
2.12 0.136501025408506
2.16 0.116136945784092
2.2 0.0580791619916756
2.24 0.0296555906534194
2.28 0.0263575315475463
2.32 -0.0124239139258862
2.36 -0.0971095462640127
2.4 -0.121564490099748
2.44 -0.11407407031705
2.48 -0.126343132307132
2.52 -0.155482349296411
2.56 -0.144666352619727
2.6 -0.101799322292209
2.64 -0.078219981243213
2.68 -0.0621405340110262
2.72 -0.0416279966011643
2.76 -0.0181059989457329
2.8 -0.00925191522886355
2.84 -0.00432749868681035
2.88 -0.000737412211795667
2.92 0.000892660580575434
2.96 0.00135364942252633
3 0.00186551331231988
3.04 0.00404647924005983
3.08 0.00615939435859522
3.12 0.0076345590253671
3.16 0.00854303780943157
3.2 0.00901978804419438
3.24 0.00906300762047372
3.28 0.00637453670303028
3.32 0.00376864864180485
3.36 0.00273754168301822
3.4 0.00677769569059213
3.44 0.0250781706223885
3.48 0.076295556810995
3.52 0.203569396088521
3.56 0.354486702630917
3.6 0.493096052358548
3.64 0.588998195404808
3.68 0.60128470013539
3.72 0.528292023887237
3.76 0.288683143444359
3.8 -0.0102032547195752
3.84 -0.290178461621205
3.88 -0.497779867301385
3.92 -0.580479931086302
3.96 -0.591498739396532
4 -0.493782631432017
4.04 -0.349158284254372
4.08 -0.21126060311993
4.12 -0.106108813546598
4.16 -0.0555873538057009
4.2 -0.0247322741585473
4.24 -0.0108137670904398
4.28 -0.00837405398488045
4.32 -0.00715631479397417
4.36 -0.00754594166452685
4.4 -0.00771830137819052
4.44 -0.00679651725416382
4.48 -0.00566453859210013
4.52 -0.00452079980944595
4.56 -0.0032895444892347
4.6 -0.00137494225054982
4.64 -2.86069698632002e-05
4.68 0.000595341902226199
4.72 0.00163422074789802
4.76 0.0043440447188914
4.8 0.0067835260803501
4.84 0.0077034744123618
4.88 0.00814630448197327
4.92 0.00857850319395463
4.96 0.00734110238651435
5 0.00264865687737862
5.04 -0.00187031722938022
5.08 -0.00416418242578705
5.12 -0.00604150972018638
5.16 -0.00786358894159395
5.2 -0.00844166210542122
5.24 -0.00691193000723918
5.28 -0.0053546947116653
5.32 -0.00472796491036812
5.36 -0.00381782107676068
5.4 -0.00150135982160767
5.44 0.0002199096294741
5.48 0.00125628399352235
5.52 0.00212987558916212
5.56 0.00289574063693483
5.6 0.0033118762075901
5.64 0.00101851470147568
5.68 -0.000337187821666417
5.72 -0.00085996727769575
5.76 -0.00118719755361478
5.8 -0.00147531305750212
5.84 -0.00177608709782363
5.88 -0.000550840826084217
5.92 0.000206577436377584
5.96 0.00050539011135695
6 0.000464187469333384
6.04 0.0004196741307775
6.08 0.000515927094966167
6.12 0.000379049219191067
6.16 0.000150719347099484
6.2 3.47693761200765e-07
6.24 0.0005199257284403
6.28 0.00162032991647722
6.32 0.00360884657129647
6.36 0.0048392377793789
6.4 0.00568577352290352
6.44 0.00600495810310047
6.48 0.00545369073127708
6.52 0.00367863771195215
6.56 -0.000297153989473966
6.6 -0.0025932219189902
6.64 -0.0040996535681188
6.68 -0.00468853519608577
6.72 -0.00471695000305773
6.76 -0.0038345737072329
6.8 -0.0018117977306247
6.84 -0.000805082730948934
6.88 0.0001661650215586
6.92 0.000900025634715966
6.96 0.00126208954801165
7 0.000961576122790616
7.04 0.000696115971853383
7.08 0.000669205871721133
7.12 -1.78025414546167e-05
7.16 -0.0011891055231293
7.2 -0.00179055007174612
7.24 -0.00193468776221078
7.28 -0.00200120282048983
7.32 -0.00201573967933657
7.36 -0.00144352996721865
7.4 -0.00072571511069935
7.44 -0.000561967492103583
7.48 -0.000484667097528783
7.52 -0.000231576152145867
7.56 -0.000147876640160884
7.6 -0.000589834060519934
7.64 -0.000793052837252617
7.68 -0.00044824881479145
7.72 0.000319682216892633
7.76 0.0010148483949403
7.8 0.00290950294584037
7.84 0.00379943832134207
7.88 0.00495138298720122
7.92 0.00546876527369023
7.96 0.00486712576821448
8 0.00374072697013617
8.04 5.87616426249162e-05
8.08 -0.00112594819317262
8.12 -0.00295638277505837
8.16 -0.00429749007647238
8.2 -0.00462406392519677
8.24 -0.00456635591884457
8.28 -0.00302859395742417
8.32 -0.00259752493972582
8.36 -0.00207830468813578
8.4 -0.00176769215613603
8.44 -0.00158006030445295
8.48 -0.0014772289432585
8.52 -0.00152561037490765
8.56 -0.00196733395569025
8.6 -0.00211611203849315
8.64 -0.00219819508492945
8.68 -0.00252889601203303
8.72 -0.00280432084885737
8.76 -0.00287241706003743
8.8 -0.00236806863298017
8.84 -0.00186477764509617
8.88 -0.00153089485441645
8.92 -0.00109119053619602
8.96 -0.00043812417425215
9 6.32476682461665e-06
9.04 -8.64810620748344e-06
9.08 -8.573072652025e-05
9.12 9.409990161655e-05
9.16 0.000201785548900533
9.2 -1.10906548798167e-05
9.24 -0.000169103887553
9.28 -5.96099998801668e-05
9.32 -8.02107776204502e-05
9.36 -0.000270110477382917
9.4 -0.000415382363523017
9.44 -0.000326557162528233
9.48 -0.000207677250728016
9.52 -0.00037798813233775
9.56 -0.000327170593664067
9.6 -4.16131224483335e-05
9.64 0.000228701702629517
9.68 0.000438480249916516
9.72 0.000845033132160699
9.76 0.0033900741642962
9.8 0.00248679184975725
9.84 0.00125261667805415
9.88 2.38317685822831e-05
9.92 -0.00136107873792447
9.96 -0.00318616962370768
10 -0.00881249729233478
};
\addlegendentry{{\scriptsize Deltas}}
\addplot [line width=1.08pt, green01270, mark=triangle*, mark size=2, mark options={solid,rotate=180}, only marks]
table {%
0 0.058752957376341
0.48 0.689289328907927
1.08 0.549836547379692
1.36 0.0573334376811977
1.52 0.0819576164881384
2.12 0.159501025408506
2.44 -0.0910740703170501
3.24 0.0320630076204733
3.68 0.62428470013539
4.32 0.015843685206028
4.92 0.0315785031939567
5.6 0.0263118762075922
5.96 0.023505390111359
6.08 0.0235159270949683
6.44 0.0290049581031026
6.96 0.0242620895480138
7.56 0.0228521233598412
7.92 0.0284687652736924
8.48 0.0215227710567437
9 0.0230063247668268
9.16 0.0232017855489027
9.28 0.022940390000122
9.48 0.0227923227492741
9.76 0.0263900741642983
};
\label{onsets}
\addplot [line width=1.08pt, red, mark=triangle*, mark size=2, mark options={solid}, only marks]
table {%
0.08 -0.000806917391717468
0.76 -0.692396439567209
1.32 0.00101672800381947
1.44 -0.0302776973247531
1.8 -0.629304646780094
2.4 -0.144564490099748
2.52 -0.178482349296411
3.36 -0.0202624583169822
3.96 -0.614498739396531
4.4 -0.0307183013781884
5.2 -0.0314416621054191
5.84 -0.0247760870978215
6.04 -0.0225803258692204
6.2 -0.0229996523062367
6.72 -0.0277169500030556
7.32 -0.0250157396793344
7.64 -0.0237930528372505
8.2 -0.0276240639251946
8.76 -0.0258724170600353
9.08 -0.0230857307265181
9.24 -0.0231691038875508
9.4 -0.0234153823635209
9.52 -0.0233779881323356
10 -0.0318124972923326
};
\label{offsets}
\end{axis}
\node [draw,fill=white] at (rel axis cs: 0.76,0.2) {\shortstack[l]{
\ref{onsets} \scriptsize Local max. \,\,\ref{offsets} Local min.}};
\end{tikzpicture}
    \vspace{-4mm}
    \caption{Proposed change-detection-based \gls{SEBB} prediction.}
    \label{fig:change_detection}
\end{figure}
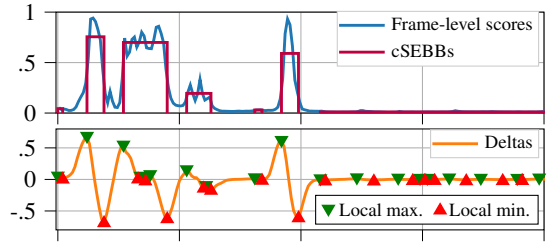

\noindent {\bf Change-detection-based \texorpdfstring{\glspl{SEBB}}{SEBBs}:}
Alternatively, we propose the following change-detection-based algorithm. First, we compute ``delta'' (i.e., change) scores by filtering $y_{n,c}$ with an ideal step filter. As different systems use different frame lengths, we perform the filtering in continuous time, interpolating $y_{n,c}$ as framewise constant. As such, for filter length $\tau_c$ (in seconds), a delta score corresponds to the difference between the average of $y_{n,c}$ in the next $\tau_c/2$ seconds and the previous $\tau_c/2$ seconds.
Now, local maxima (resp. minima) of the delta scores become tentative onsets (resp. offsets), forming tentative events (resp. gaps) between each onset (resp. offset) and the next offset (resp. onset).
Then, as some tentative gaps may be due to only small spurious variations of $y_{n,c}$, we employ the following merging strategy. For every tentative gap, we compare its lowest $y_{n,c}$ with the highest $y_{n,c}$ in the tentative events immediately preceding and following it. If the comparisons fall under a predefined merge threshold $\gamma_c$, the tentative offset and onset around the gap are removed (i.e., the tentative preceding event, the following event and the gap between them are merged into the same event). We test $\gamma_c$ both as threshold on the difference (i.e., absolute threshold) and ratio (i.e., relative threshold) between scores.
Finally, we form a predicted \textit{cSEBB} of class $c$ from each remaining onset as $\hat{t}_{\text{on}}$, the following remaining offset as $\hat{t}_{\text{off}}$, and the average of $y_{n,c}$ between $\hat{t}_{\text{on}}$ and $\hat{t}_{\text{off}}$ as $\overline{y}$. We then have the class-dependent filter length $\tau_c$ and threshold $\gamma_c$ (either absolute or relative), as hyperparameters to tune on a validation set. Fig.~\ref{fig:change_detection} shows the process using a relative $\gamma=3$ (i.e., checking if neighboring tentative events' maxima fall below 3 times a gap's mimimum).

\noindent {\bf Hybrid \texorpdfstring{\glspl{SEBB}}{SEBBs}:}
We further propose a hybrid of the two previous approaches, where we predict a set of \textit{hSEBB}s as follows. We first predict tSEBBs and select those above a certain confidence $\lambda_{c,\text{hyb}}$. These are then complemented by cSEBBs, discarding any cSEBB that overlaps with selected tSEBBs. Here, the predicted cSEBBs may find additional low-confidence \glspl{SEBB}. We can tune the following class-level hyperparameters on a validation set: median filter lengths and $\lambda_{c,\text{ext}}$ for tSEBB prediction, $\{\tau_{c},\gamma_{c}\}$ for cSEBB prediction, and $\lambda_{c,\text{hyb}}$.

\vspace{-.1cm}
\section{Experiments}
\vspace{-.15cm}

\begin{figure*}[t]
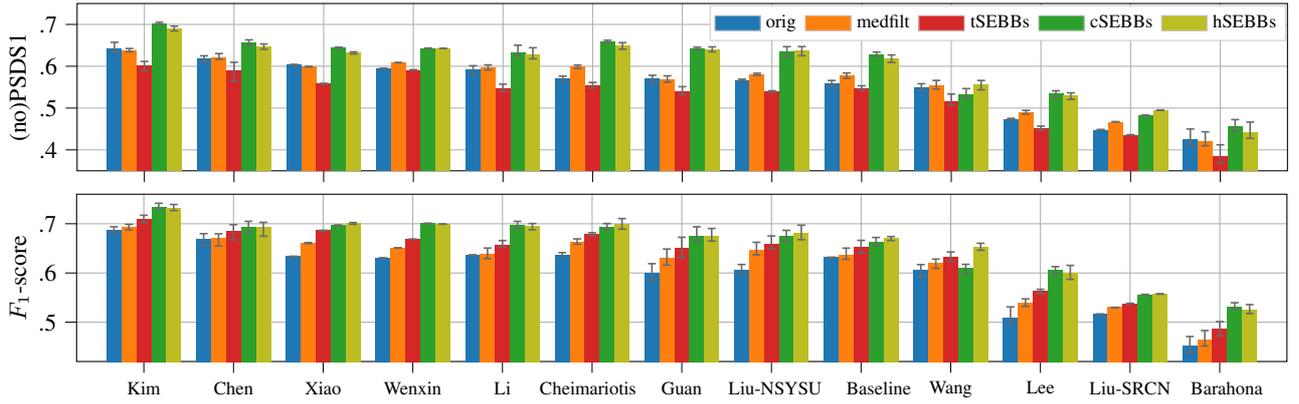

    \centering
    \setlength\figureheight{3.8cm}
    \setlength\figurewidth{17.5cm}
    \input{figures/psds_cv.tex}\\
    \vspace{-1mm}
    \input{figures/cbf_cv.tex}
    \vspace{-7mm}
    \caption{Results on public evaluation set using 5-fold cross-validation for hyper-parameter tuning.}
    \label{fig:results_cv}
    \vspace{-4mm}
\end{figure*}

We evaluate our proposed methods on DCASE 2023 Challenge Task 4 submissions~\cite{ebbers_2023_8248775} which target \gls{SED} in domestic environments using the DESED dataset~\cite{DESED}.
A system submission consisted of three timestamped frame-level score series for the evaluation set generated corresponding to three (independent) system training runs.
Additionally, if a proposed system included any frame-level post-processing, participants were asked to also provide ``raw'' frame-level scores before post-processing. %

We run our evaluations on the public portion of the evaluation set for which ground-truth annotations are publicly available.
To not apply our methods on top of other post-processing schemes, we only consider teams that provided raw scores, i.e., 13 teams (counting the baseline).
For each team, we only report the system that performed best in terms of the challenge \gls{PSDS}1 metric, %
i.e., Kim-2~\cite{Kim2023}, Chen-2~\cite{Chen2023}, Xiao-4~\cite{Xiao2023}, Wenxin-6~\cite{Wenxin2023}, Li-6~\cite{Li2023}, Cheimariotis-1~\cite{Cheimariotis2023}, Guan-3~\cite{Guan2023}, Liu-NSYSU-7~\cite{LiuNSYSU2023}, Baseline-2~\cite{baseline2023beats}, Wang-1~\cite{Wang2023}, Lee-1~\cite{Lee2023}, Liu-SRCN-4~\cite{LiuSRCN2023a}, and Barahona-2~\cite{Barahona2023}.

As evaluation metrics, we use two metrics from the challenge, i.e., 1) \gls{PSDS}1, i.e., PSDS with ${\rho_\text{DTC}=\rho_\text{GTC}=0.7}$ and an across-class standard-deviation penalty weight of $\alpha_\text{ST}=1$, and 2) collar-based $F_1$-score~\cite{mesaros2016metrics}, henceforth simply referred to as $F_1$, with a \SI{200}{ms} onset and max(\SI{200}{ms}, \SI{20}{\%} of ground truth event length) offset collar.
We are not considering the \gls{PSDS}2 metric, as it is more tuned as an audio tagging metric than an \gls{SED} metric~\cite{nam2021heavily}.

As previously mentioned, \gls{PSDS} considers only best-case decision thresholds, i.e., thresholds leading to less \glspl{TP} than a higher threshold are discarded from the ROC curves.
Whether an operating point is best-base or not, however, can only be determined by evaluation w.r.t. ground truth.
This is then in square contradiction with any practical scenario where this oracle information would be, of course, inaccessible.
Therefore, when using legacy event prediction (i.e., frame-level thresholding followed by merging), we also report \textit{\gls{noPSDS}}, i.e., \gls{PSDS} without best-case selection, that is, the normalized area under the (possibly non-monotonic) PSD-ROC.
In order to simply limit the descent of the PSD-ROC, however, we pre-tune minimum thresholds $\lambda_{c,\text{noPSDS}}$, for each class $c$, which give the maximal number of \glspl{TP} (\#TP) on a validation dataset (i.e., below which \#TP only decreases).

For each system, we report performance using legacy event prediction in conjunction with the original submission post-processing (orig) and common median filter post-processing (medfilt), and for SEBB-level thresholding in conjunction with our three proposed post-processing methods which output tSEBBs, cSEBBs, and hSEBBs, respectively.
For each system and event class, the following hyper-parameters are to be tuned on a validation set:
\begin{compactitem}
\item orig: $\lambda_{c,\text{noPSDS}}$,
\item medfilt: median filter length, $\lambda_{c,\text{noPSDS}}$,
\item tSEBBs, cSEBBs, hSEBBs: see lists in Sec.~\ref{ssec:SEBBalgos},
\end{compactitem}
plus, for each method, a decision threshold $\lambda_{c,F}$ for $F_1$ evaluation.
Different hyperparameter sets are tuned for PSDS and $F_1$ evaluation, respectively.  
Optimal thresholds can be efficiently tuned using sed\_scores\_eval~\cite{ebbers2022threshold}.
Median filter lengths are chosen out of $\{\SI{0}{s}\,\text{(no filter)}, \SI{.2}{s},\dots,\SI{2}{s}\}$.
$(\tau_c,\gamma_c)$ is chosen out of $\{\SI{.32}{s},\SI{.48}{s},\SI{.64}{s}\}\times\{.15\,\text{abs.},.2\,\text{abs.},.3\,\text{abs.},1.5\,\text{rel.},2\,\text{rel.},3\,\text{rel.}\}$.
Hyperparameters are tuned to maximize the respective metric with the following exception.
For hSEBBs, for simplicity, we don't tune all parameters. Instead, we adopt tSEBB-related parameters from stand-alone tSEBBs and cSEBB-related parameters from stand-alone cSEBBs and optimize only $\lambda_{c,\text{hyb}}$.

\begin{figure}[t]
    \centering
    \setlength\figureheight{3.5cm}
    \setlength\figurewidth{8.40cm}
    \input{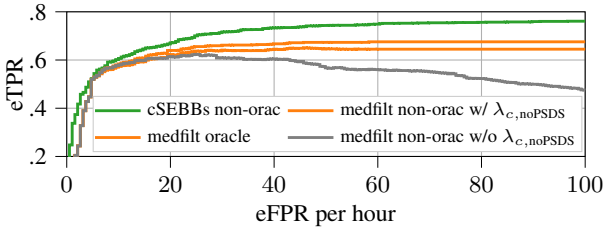}
    \vspace{-7mm}
    \caption{PSD-ROCs for Kim with different post-processing.}
    \label{fig:psdrocs}
\end{figure}

Without access to outputs on the challenge's validation set, we instead report noPSDS1 (using $\lambda_{c,\text{noPSDS}}$ for legacy event prediction) and $F_1$ for a 5-fold cross-validation on the evaluation set outputs, where predictions for each fold are generated using hyper-parameters tuned on the four other folds, in Fig.~\ref{fig:results_cv}. For each condition, we show the mean, lowest and highest score over the three provided runs. For legacy event prediction, we also show the (higher) PSDS1 using a lighter color.
Interestingly, we see that most systems would already perform better with legacy event prediction if they just traded their current post-processing for a straightforward class-specific median filter post-processing. We only base this conclusion on the (no)PSDS1 scores though, as most systems likely tuned hyperparameters only once for PSDS1, penalizing their $F_1$ score, while we tune separate hyperparameter sets for each metric.
PSDS1 performance degrades with tSEBBs, which suggests that indeed the same threshold cannot correctly predict the extents of events that have different detection confidences.
For $F_1$-score, 
which evaluates a single operating point, it 
however is clearly beneficial to have different extent detection and event prediction thresholds.
For the more sophisticated cSEBBs and hSEBBs, they overall substantially outperform the other methods.
cSEBBs improves over median filtering for all systems but Wang \cite{Wang2023}, achieving an average gain of $\SI{4.1}{\%pt.}$ for PSDS1 and $\SI{3.4}{\%pt.}$ for $F_1$.
They boost the winning system's PSDS1 from .644 to .703, and $F_1$-score from .688 to .734 setting the state of the art on this particular setup. 
At the same time, hSEBBs outscores median filtering for all systems, albeit only improving over cSEBBs for a select few. That hSEBBs detoriate performance over cSEBBs for most systems can be explained by poorer generalization of the increased number of hyper-parameters. %

In Fig.~\ref{fig:psdrocs}, we further see the expected benefits of cSEBBs on the Kim-2 system, when comparing the corresponding PSD-ROC curve with curves for the legacy event prediction with median filter post-processing. First, it can be seen how the oracle modification of PSDS distorts the intuition behind the AUC-like component of the PSDS, substantially diverging from the non-oracle curve, which is partially mitigated by the addition of $\lambda_{c,\text{noPSDS}}$.
Further we can see that the inflated oracle modification still fails to close the gap with the cSEBBs' PSD-ROC curve, ending up lower in every operating range.

Finally, to evaluate our method in the proper challenge setting, i.e., tuning our hyperparameters on validation set output scores, we contacted participants and asked whether they could share these\footnote{We would like to thank all teams who responded to our request.}.
We received raw validation scores from Xiao~\cite{Xiao2023}, Li~\cite{Li2023}, Barahona~\cite{Barahona2023} and the baseline~\cite{baseline2023beats}, and postprocessed validation scores from Kim~\cite{Kim2023}.
We then optimize hyperparameter sets on that data before scoring on the full evaluation set. Note that only having postprocessed validation scores for Kim means the cSEBB post-processing effectively differ at validation (i.e., with original post-processing added) and evaluation (i.e., without).
Our method, again, significantly improves performance for all systems and we achieve new challenge state of the art performances of .686 PSDS (by Kim~\cite{Kim2023}) and .706 $F_1$ (by Xiao~\cite{Xiao2023}), with Kim's performance likely being hurt by the validation/evaluation cSEBB mismatch.

\vspace{-.2cm}
\section{Conclusions}
\vspace{-.2cm}
In this work, we demonstrated how the commonly used frame-level thresholding for SED results in a harmful coupling of event extent and confidence prediction.
As solution, we introduced \acrfullpl{SEBB} as a new general SED output format, which also overcomes the ill-definition of recent event-based evaluation metrics.
We further proposed a change-detection-based algorithm to infer SEBBs from frame-level model outputs.
Our experiments showed that our proposed method allows for substantially improved performance for a large range of systems and sets a new state of the art on the DCASE 2023 Challenge Task 4a benchmark.

\balance
\bibliographystyle{IEEEtran}
\bibliography{mybib}

\end{document}